# Structural colors and enhanced resolution at the nanoscale:

## Local structuring of phase-change materials using focused ion beam

Daniel T. Yimam*, Minpeng Liang, Jianting Ye, Bart J. Kooi*

Zernike Institute for Advanced Materials, University of Groningen, Nijenborgh 4, 9747 AG Groningen, The Netherlands

## Abstract

In the past few years, phase-change materials have become increasingly important in nanophotonics and optoelectronics. The advantages of sizeable optical contrast between phases and the additional degree of freedom from phase switching have been the driving force. From multi-level reflectance to dynamic nanoprinting and structural colors, phase-change materials have achieved outstanding results with prospects for real-world applications. The local crystallization/amorphization of phase-change materials and the corresponding reflectance tunning by the crystallized/amorphized region size have potential applications for future dynamic display devices. Although the resolution is much higher than current display devices, the pixel sizes in those devices are limited by the locally switchable structure size. Here, we reduce the spot sizes further by using ion beams instead of laser beams and dramatically increase the pixel density, demonstrating the capability of having superior resolution. In addition, the power to sputter away materials can be utilized in creating nanostructures with relative height differences and local contrast. Our experiment focuses on one archetypal phase-change material, $Sb_2Se_3$, prepared by pulsed-laser deposition on a reflective gold substrate. We demonstrate that we can produce structural colors and achieve reflectance tunning by focused ion beam milling/sputtering of phase change materials at the nanoscale. Furthermore, we show that the local structuring of phase-change materials by focused ion beam can be used to produce high pixel density display devices with superior resolutions.

# Introduction

Optical materials with phase-change transitions have been heavily studied in nanophotonics and optoelectronics for potential applications in data storage and display devices.[1] Researchers are heavily invested in generating structural color to curtail power consumption and pixel density scaling problems[2] in conventional display devices. The production of vivid and wide gamut structural colors has been reported using Fabry-Perot cavity[3,4] and metal metasufaces[5–7]. The successful adaptation of the strong interference effect, where a lossy thin film is coated on a highly reflective surface, led to significant advances in nanophotonics in recent years.[8–11] One attractive feature, especially for future display device applications, is how robust the structures are to changes in the angle of incidence.

Even though outstanding results have been reported, moving away from a static system towards dynamically reconfigurable modules seems complicated. Some progress towards dynamically reconfigurable optical devices involves optical properties tunning active materials throughout chemical reactions[12,13], liquid crystals[14], and structural changes of phase-change materials (PCMs).[15–19] PCMs quickly became the focus for applications in dynamically reconfigurable photonics, including structural color production and future display device applications.[1] The optical contrast between different phases of PCMs and the capability of switching between phases suit the PCM to reflectance-based applications. In addition, the ability to produce partial structural states and tune the reflectance states seems promising for moving forward to a dynamic display device.[15,20,21] Multiple research works have already shown the partial crystallization/amorphization of PCMs by using laser pulses,[15,17,20,22] electrical triggering,[16,18,23] and ion beams.[24] The associated structural changes, and thus reflectance changes, produced multiple states and optical contrasts with significant application potential for future optoelectronics.

In this work, we demonstrate the production of structural colors locally at the surface of reflective heterostructures. We employed the milling/sputtering power of a focused ion beam (FIB) to locally nanostructure an optical reflectance device consisting of a reflective gold substrate and an $Sb_2Se_3$ film on top. Although much studied for solar cell applications,[25–28] $Sb_2Se_3$ is one of the low-loss phase change materials that recently attracted growing interest due to its optical properties in both the as-deposited amorphous and the crystalline phases with large optical contrast between them[19,29,30]. Furthermore, the low extinction coefficient in the visible and near-infrared spectral ranges, the high index of refraction contrast between the two phases, and the wide bandgap are promising for coupling the material's phase-switching properties with optoelectronic devices.[15,31] Here, we experimentally demonstrate local thickness variations and the associated reflectance tuning for producing super high-resolution nanoprints with diffraction-limited pixel resolutions. We also show that FIB can create various contrasts throughout the thickness of the active material, adding another degree of freedom for acquiring all accessible structural colors. Our work offers an innovative approach promising for future display devices with nanostructured surfaces, which could be coupled with partial switching of phase change thin films.

**Experimental**

$Sb_2Se_3$ thin films were pulsed laser deposited (PLD) on various substrates. A laser fluence of 1 J cm$^{-2}$, a processing gas (Ar) of 0.12 mBar, and a repetition rate of 1 Hz were used for the depositions. The PLD system has a KrF excimer laser operating at 248 nm wavelength. $Sb_2Se_3$ thin films were deposited on $Si_3N_4$ membranes for elemental composition analysis and local imaging by transmission electron microscope (TEM). A TEM (JEOL 2010) operating at 200 kV accelerating voltage and an energy dispersive x-ray (EDX) detector was used to analyze the as-deposited and crystalline $Sb_2Se_3$ thin films.

Spectroscopic ellipsometry (J. Woollam UV-VIS) was used to extract the optical constants of $Sb_2Se_3$ thin films in the spectrum range of 300 – 1700 nm. For ellipsometry, multiple samples of $Sb_2Se_3$ with varying thicknesses were deposited on thermal $SiO_2$ substrates. A spectroscopic scan was done within the spectral range. For both amorphous and crystalline $Sb_2Se_3$ samples, measurements of $\psi$ and $\Delta$ were collected from multiple angles of incidence (65º, 70º, 75º) for increased data fitting accuracy and reduced parameter correlations. The data fitting/analysis of the measured ellipsometry date was done with the WVASE software. A heterostructure model was built, and the Tauc – Lorentz optical oscillator was used to model the optical properties of the $Sb_2Se_3$ thin films. During data fitting, the Tauc – Lorentz oscillator parameters and the film thickness were allowed to vary, and the refractive index (n) and extinction coefficient (k) were extracted. The crystallization temperature ($T_x$) of the as-deposited $Sb_2Se_3$ thin films was extracted from dynamic ellipsometry (DE) measurements. A heating stage (HTC-100) was attached to the variable-angle spectroscopic ellipsometry (VASE) setup, and the TempRampVase software controller was used for temperature ramping. A 70º incidence angle and a heating rate of 5 ºC min$^{-1}$ were used for all samples.

Reflectance profiles of multiple reflective heterostructure devices have been calculated by employing the transfer-matrix algorithm. The ellipsometry-extracted values for the optical constants of $Sb_2Se_3$ were used for the calculations. $Sb_2Se_3$ thin films with varying thicknesses were deposited on a reflective gold (Au) layer to produce reflective devices. A 100 nm thick Au layer was first deposited on a Si substrate using an e-beam evaporator (Temescal FC2000). A Ti layer, 5 nm thick, was first deposited before the gold layer to increase its adhesion. To avoid intermixing between the Au layer and the $Sb_2Se_3$ thin film, a 10 nm spacer $LaAlO_x$ (LAO) layer was deposited first on top of the Au layer. In some samples, a 10 nm capping LAO layer was deposited on top of the $Sb_2Se_3$ layers.

The local structuring of the $Sb_2Se_3$ was done by a focused ion beam (FIB of FEI Helios G4 CX Dual Beam system) operating at 30 kV accelerating voltage. The grayscale value of an image was initially calibrated to the milling power of the $Ga^+$ ion beam. Then, an image was loaded, and the corresponding pattern was created by ion beam milling (sputtering away materials). Finally, the morphologies of the completed structures were imaged by the scanning electron microscope (SEM, Helios G4 CX) and by atomic force microscopy (AFM, Dimension Icon, Bruker). The reflectance spectra of the locally nanostructured areas were measured using a home-built system to collect intensity from <10 μm regions. The setup uses white light and is focused by a 100x objective with a numerical aperture of 0.9. The collected light intensity was analyzed by a spectrum analyzer and normalized to a reference signal.

## Results and discussion

Thin films of $Sb_2Se_3$ were deposited using PLD on a $Si_3N_4$ membrane. For the structural analysis the as deposited amorphous and crystalline phases of $Sb_2Se_3$ thin films, we used transmission electron microscopy. Figure 1a – c show the TEM images of global (a), local (b) morphologies, and the selected area electron diffraction (SAED) patterns (c), respectively. The as-deposited films are globally amorphous. The amorphous nature of the as-deposited films is confirmed by SAED, showing diffusive rings as shown in the inset of Fig. 1a. In Fig. 1b and c, higher resolution images focused on more local areas show the crystalline sample morphology and a diffraction pattern from the SAED shows well-defined spots (Fig 1c). $Sb_2Se_3$ crystallizes in an orthorhombic crystal structure (space group Pnma). The overall composition of the deposited film has, on average (in at.%), 41 Sb and 59 Se, confirming the production of rather exact stoichiometry.

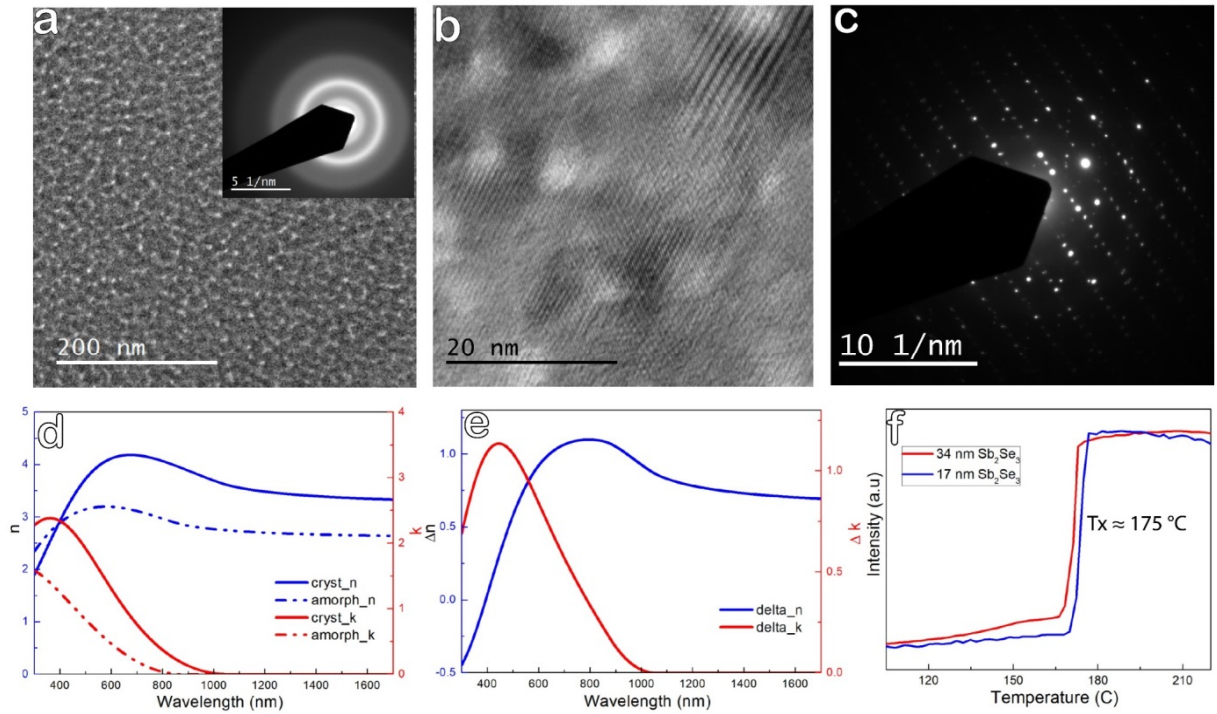

**Figure 1**. Structural and optical analysis and measurement of pulsed laser deposited $Sb_2Se$ thin films. (a) TEM characterization of an as-deposited phase of $Sb_2Se_3$ thin film on $Si_3N_4$ membrane. The as-deposited phase is entirely amorphous, as confirmed by the FFT pattern in the inset. (b, c) TEM image and associated SAED pattern of a crystalline $Sb_2Se_3$ thin film. The as-deposited phase was thermally annealed at 250 °C for 5 min to induce crystallization. (d) Spectroscopic ellipsometry measurements of as-deposited and annealed $Sb_2Se_3$ thin films deposited on thermally grown $SiO_2$ substrates were used to extract refractive index (n) and extinction coefficients (k) for a wide spectral range of 300 – 1700 nm. (d) A significant difference in n and k values between the as-deposited amorphous and crystalline samples is seen. (f) The precise crystallization temperature (Tx) of the as-deposited $Sb_2Se_3$ thin films is extracted from dynamic ellipsometer measurements.

For the characterization of optical responses in different phases, $Sb_2Se_3$ films with varying thicknesses were deposited on thermal $SiO_2$/Si substrates for spectroscopic and dynamic ellipsometry characterizations. Figure 1d presents the optical constants (index of refraction *n* and extinction coefficient *k*) of the as-deposited and crystalline phases of $Sb_2Se_3$ thin films. The characterizations start with a spectroscopic scan of the as-deposited phase in the spectral range of 300 – 1700 nm, followed by the dynamic ellipsometry where the light intensity was continuously probed while the as-deposited films are annealed with a constant heating rate of 5 °C/min. Since the optical properties of the as-deposited and crystalline phases are different, the light intensity will dramatically change precisely at the crystallization temperature Tx of the sample. Therefore, accurate crystallization temperature values can be extracted from the first derivative of the intensity plot. As shown in Fig. 1f, the crystallization temperature is $T_x \approx 175$ °C. Subsequently, we use spectroscopic ellipsometry for the crystalline phase. A model describing the sample's heterostructure must be constructed to extract the optical constants from the measured ellipsometry parameters $\psi$ and $\Delta$. The Tauc-Lorentz oscillator model can explain the optical constants of the amorphous (as-deposited) and crystalline (above $T_x$) phases of $Sb_2Se_3$ films. The dielectric constant values in Fig. 1d also correspond well with previously reported results[32,33]. The index of refraction ($n$) value changes between the phases has maximum values in the visible frequency spectrum, as seen in Fig. 1e. This large change can be directly related to a significant reflectance change between amorphous and crystalline phases, attractive for dynamic display devices.

After knowing the ellipsometry optical constants of $Sb_2Se_3$, we can design a heterostructure display device, and the reflectance spectrum can be simulated based on the transfer matrix (TMM) formalism. Fig. 2a shows the reflective device design schematics where the active $Sb_2Se_3$ material was pulsed laser deposited on a reflective gold substrate. Spacer LAO layers were deposited on the bottom and top of the active material for capping to avoid intermixing between $Sb_2Se_3$ with the gold substrate and evaporation during crystallization at high temperatures. It is well known that film thicknesses play a crucial role in tuning the reflectance spectra in the strong interference formalism where an absorbing layer is coated on a highly reflective surface.[8] Fig. 2b shows the experimental and simulated reflectance spectra for the $Sb_2Se_3$ films with thicknesses of 4 – 24 nm. The measurements and simulations are given for the as-deposited amorphous and crystalline phases of the $Sb_2Se_3$ layer. An excellent agreement is reached by comparing the calculated reflectance spectra with the measured values. Different structural colors, corresponding to $Sb_2Se_3$ film thickness variations, are shown in Fig. 2c. Multiple shades of structural colors are produced for the as-deposited phase. Another set of variable colors is also made after annealing the samples for 5 min at 250 °C. For comparison, we show, in Fig. 2d, the CIE 1931 color space and the respective color coordinates of the reflectance spectra in (b). A color-matching function was used to calculate the x and y chromatic coordinates from the reflectance spectra. The color map shows that both as-deposited amorphous and crystalline phases of the $Sb_2Se_3$ layer have sizeable RGB gamut coverages. And the amorphous – crystalline phase transition has significant color differences, especially in the two ends of the spectral range, namely towards both blue and red regions,

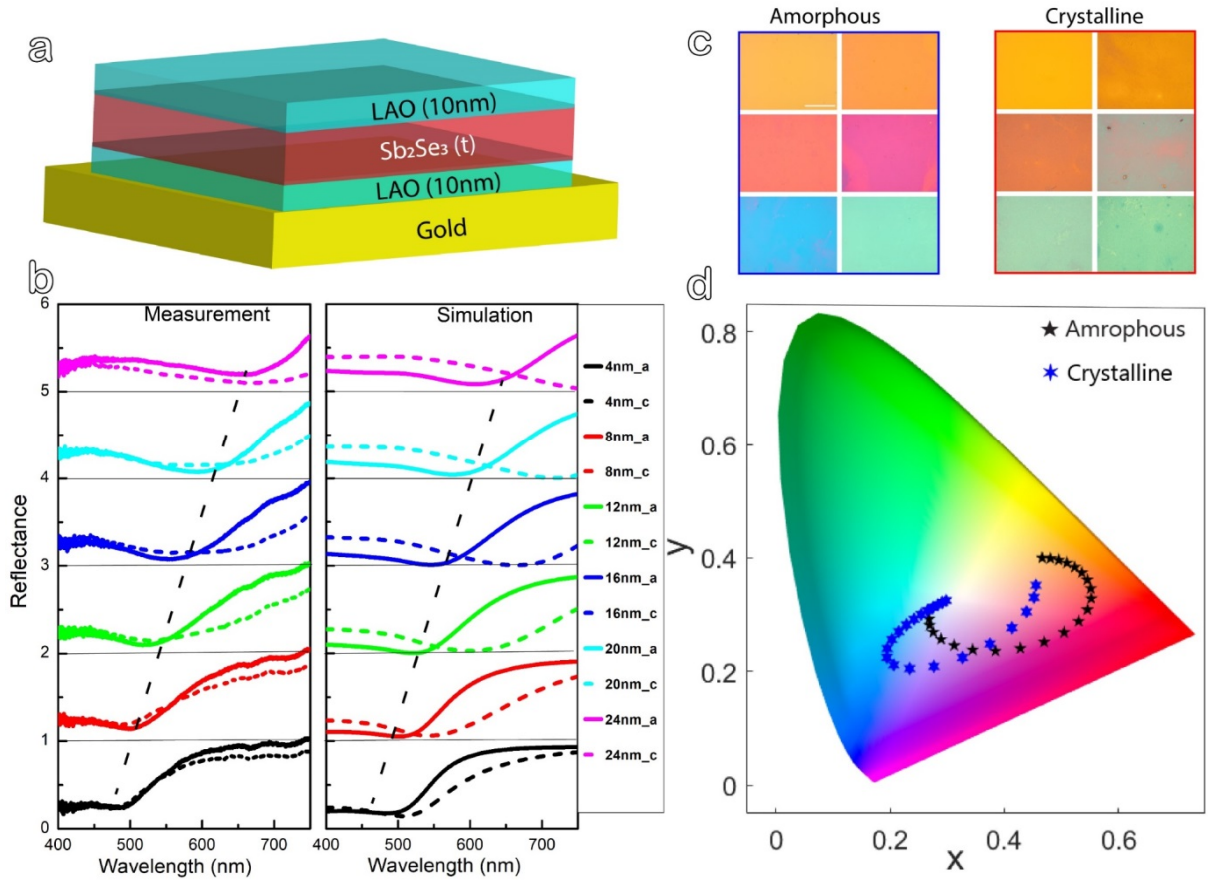

**Figure 2**. A reflective heterostructure design based on $Sb_2Se_3$ thin films on reflective gold substrates. (a) The schematics of the heterostructure stack comprise an active $Sb_2Se_3$, a reflective gold substrate, a capping, and thin spacer LAO layers. (b) Measured and simulated normal incidence angle reflectivity spectra of the heterostructure device for varying the $Sb_2Se_3$ film thickness. Both measured and simulated reflectivity spectra contain results for the as-deposited amorphous and crystalline phases of the $Sb_2Se_3$ layer. Structural colors produced by the reflective heterostructure design are presented in (c). The made colors are for the as-deposited and crystalline phases of $Sb_2Se_3$, with thickness varying from 4 – 24 nm. All images' dimensions are the same, and the scale bar is 200 μm. (d) The corresponding color coordinates for the reflectance values in (b) are plotted on a chromatic diagram for both the as-deposited amorphous and crystalline thin films of $Sb_2Se_3$.

In the strong interference formalism, where a lossy thin film is coated on a reflective surface, a slight thickness change leads to a significant shift in the reflectance spectra. Controlling/changing film thickness thus can produce all accessible structural colors. In Fig. 2, we showed how a slight change in the thickness of the active $Sb_2Se_3$ thin film layer, prepared on a reflective gold substrate, correlates to the production of different structural colors. The produced reflective heterostructure devices are smooth (due to reduced thickness values), and the reflectance is uniform across the device surface. However, additional steps are needed to produce local contrast on the surface. We change the phase of PCM by inducing partial crystallization/amorphization of active layers. By varying the local ordering of the active material with electronic signals or laser pulses, the sub-diffraction resolution of images can be nanoprinted. Although, for the contrast, the dynamic range is limited by the maximum reflectance change of the amorphous and crystalline phases of the active PCMs.

To demonstrate the possibility of reaching high local contrast, we employ the large reflectance tunning from slight thickness changes of the active material. The locally controlled thickness variation on a thin film surface utilized the focused ion beam (FIB). The schematic of the FIB-device interaction is illustrated in Fig. 3a. A heterostructure reflective device was produced by depositing an about 26 nm thick $Sb_2Se_3$ thin film on a reflective gold substrate. A Ga+ ion beam, accelerated to 30 kV, is focused onto the surface of the device. The ion beam is used to locally "sputter away" material, and the amount of removed material depends on the beam intensity and dwelling time. Therefore, calibration is needed to precisely control the FIB-device interaction and the material removal. First, a grayscale image with 25 squares of varying grayscale values (from 10 to 250 on an 8-bit grayscale) was used to pattern the surface of the device (see supplementary information (SI) Fig. S1a). Then, by carefully experimenting with different milling parameters of the ion beam, optimum values for local structuring of the reflective device surface were acquired. More importantly, the beam current, the milling depth, and the beam dwell time are calibrated to allow controlled milling to sub-nm accuracy. Fig. 3b shows an SEM image where the contrast is created from the collected backscattered elections (BSEs) signal. Each square has a lateral dimension of $10\times10$ μm$^2$. The height difference of individual squares can be seen from the SEM image. Since the BSEs intensity scales with the atomic number (Z) of elements, i.e., higher *Z* elements scatter electrons more than low *Z* elements, the contrast among individual squares in the SEM image indicates the depth profile. The contrast gradient, darker from the top and lighter at the bottom, is due to how close the imaged surface is to the gold (high *Z*) substrate.

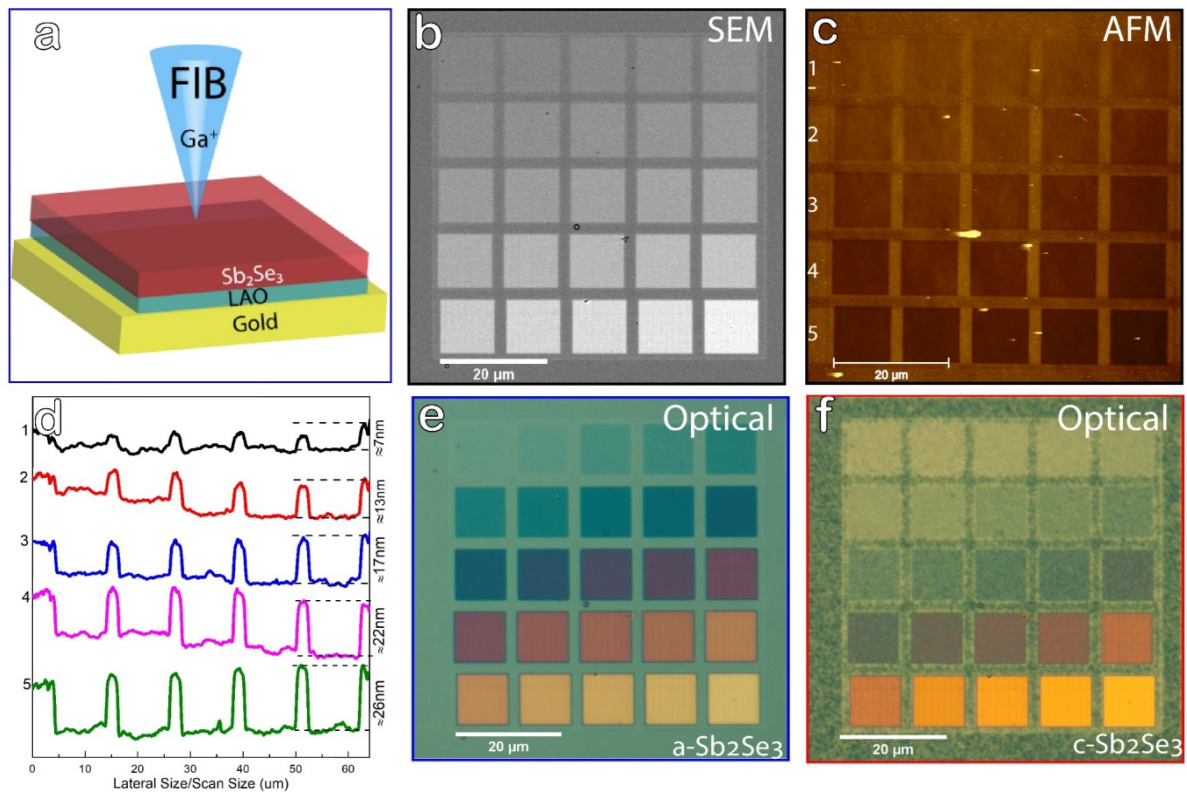

**Figure 3**. Thin film thickness vs. FIB milling/sputtering power calibration with digital grayscale values. (a) A schematic of the reflective heterostructure device produced by pulsed laser deposition of $Sb_2Se_3$ thin film on a reflective gold substrate. A Ga+ ion beam is used to create a nanostructured metasurface on the surface of the reflective heterostructure device. (b) An SEM image obtained from backscattered electrons (BSEs). The contrast between individual squares is due to how close the squares are to the bottom gold substrate. (c) An AFM scan of the square structures and (d) the line profiles of individual rows showing the local height differences. The optical images of the structured squares with multiple structural colors have been produced due to local thickness variations of the $Sb_2Se_3$ thin film for (e) the as-deposited and (f) crystalline thin film.

The precise control of local height is characterized further by atomic force microscopy (AFM). The morphologies of the patterned squares have been investigated using AFM, and the collected images are presented in Fig. 3c. The local height difference of the individual squares is seen from the AFM line profiling. From rows one to five, the local contrast changes from lighter to darker, indicating deeper regions. The line profiles over each row are also given in Fig. 3d. The height change from the top surface is shown by the steps present in the line scan. Squares in the top rows show a smaller height difference than squares in the bottom rows (see supplementary information (SI), Fig. S1b). From the SEM image in Fig. 3b and AFM image and line profiles in Fig. 3c and 3d, the visual changes can be quantitatively correlated to the height based on the grayscale values in the individual squares. As shown in Fig. 2, the local height contrasts correspond to a reflectance change and, thus, to the variable structural colors. The as-prepared structures are optically imaged and are presented in Fig. 3e and 3f for both the as-deposited and crystalline phases of the $Sb_2Se_3$ film. A wide variety of colors throughout the film thickness is created from the height variations. The reflectance profiles of individual squares seen in Fig. 3e and f have been measured and compared with calculated values. The results are presented in the supplementary information (SI) Fig. S2.

Once the milling power of the FIB was calibrated, i.e., the grayscale value to thickness conversion for a given milling/sputtering parameters was obtained, we could nanoprint any image onto the surface of the reflective display device. A heterostructure reflective device similar to what is reported in Fig. 3 was used for the proof of concept. The FIB software first loaded a digital image to produce local height variation in nanosized pixel areas. Then, two famous paintings, *Girl with a Pearl Earring* by Johannes Vermeer and *Starry Night* by Vincent van Gogh, were nanoprinted on the device. The original digital images are presented in Fig. 4a and b. Once the digital pictures were loaded and optimum beam parameters were set, nanostructures of different lateral dimensions were produced. Figure 4c and d show the BSE-SEM images of the drawn nanostructures. The contrast from the SEM images perfectly resembles the original digital images used for patterning. As discussed earlier, the BSEs signal depends on how deep a region is and how close it is to the gold substrate. In Fig. 4c and 4d, regions closer to the gold substrate are brighter, and regions close to the surface are darker.

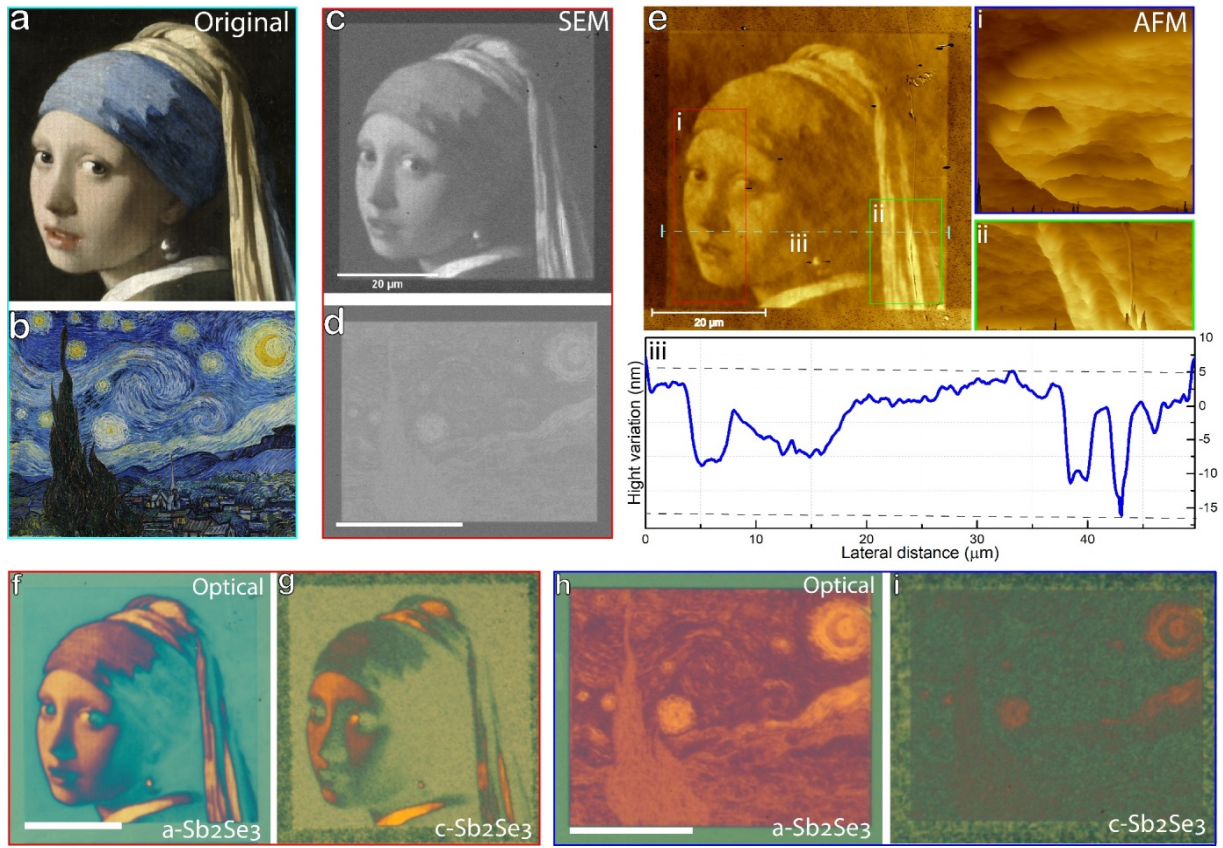

**Figure 4**. Super-resolution nanoprinting of a grayscale image using a focused ion beam (FIB). (a) The original grayscale image, *Girl with a Pearl Earring* by Johannes Vermeer and (b) *Starry Night* by Vincent van Gogh, to be printed on the Sb$_2$Se$_3$ surface. The milling rate of the FIB was calibrated beforehand so that the sputtering rate is directly related to the grayscale value of each pixel. In (c) and (d), backscattered electrons SEM images of the fabricated metasurfaces are presented. The backscattered electrons provide depth profiles. (e) A large area AFM scan of the metasurface after milling the surface. Here whiter regions are deeper, and darker are close to the surface. In (i) and (ii), the 3D views of some of the regions in the AFM scan are presented to show the depth profile created by the FIB. In (iii), a line profile on the surface of the metastructure is shown. The optical images of the reflective metasurfaces are presented in (e) and (f). The local height difference between individual pixels created high contrast. In addition, phase switching produced another degree of freedom for contrast creation.

To better visualize the local height difference and the origin of contrast in the structured images, we investigated the morphologies by AFM. In Fig. 4c, an AFM image showing the patterned region, a 3D view of some local structures (i and ii), and a line profile across the structured region (iii) are presented. The contrast from the AFM image, with local height changes, perfectly resembles the original digital image. However, it is better to have a zoomed-in view of the structured surface to fully unravel the source of the contrast in the SEM and AFM images and, ultimately, in the optical images. Therefore, in Fig. 3e (i) and (ii), a closer look and an exaggerated 3D view of the morphology are given for better visualization. The red and green boxes represent the corresponding regions in the AFM image. The 3D views of the structure show the nanosize local height variability. The terrain is comprised of higher peaks and low valleys correlated with the relative grayscale value changes of the original digital image. This change is better visualized in the line profile of the structure presented in (iii). As can be seen, the line profile shows the morphology change with a depth variability of ≈20 nm.

The local height changes seen in the SEM and AFM images and the corresponding contrast should also translate to the formation of different local structural colors from the thickness changes. Thus, the optical images should produce vivid resolutions and contrast from the reduced individual pixel sizes. Fig. 4f – 4i shows optical images of the fabricated structures for the as-deposited and crystalline phases of the $Sb_2Se_3$ thin film layer. Indeed, from the optical images, super-resolution images are visible, with a relatively small region producing extremely large DPI values. For example, the original uploaded digital image, Fig. 4a, is close to 2k x 2k pixel size, and the fabricated region in Fig. 4f is only 40 x 40 $\mu m^2$. This corresponds to a significant pixels-per-inch (PPI) value of ≈127,000. The high PPI values are achieved because of the relatively small probe size of the ion beam for an accelerating voltage of 30 kV and beam current of 0.23 nA. This is a crucial point which is discussed in more detail below.

The use of FIB for nanostructuring a reflective device's surface has numerous advantages. For one, the small probe size means we can produce height differences with nanosized lateral dimensions producing vivid structural colors with very high PPI values. However, the functionality and advantage of using FIB extend beyond what is reported in Fig. 4. The milling/sputtering power of the FIB, correlated to the grayscale values uploaded, is linearly related to the milling depth or the local height profiles created. While keeping the relative height differences constant, we can produce structured surfaces and various contrasts throughout the film thickness by varying the absolute depth. Fig. 5a shows the schematic description of how the structures could be made at different depth levels throughout the film thickness. Fig. 5b shows variable contrasts of a nanostructured surface created at different depth levels of the $Sb_2Se_3$ thin film layer. The nanostructures and the contrasts in Fig. 5b(i). are produced by increasing the milling/sputtering power of the FIB by a small amount starting from the top. Varying shades of contrasts with superb resolutions can be seen by accessing all different thickness ranges of the active material. This is also another crucial point discussed in more detail below.

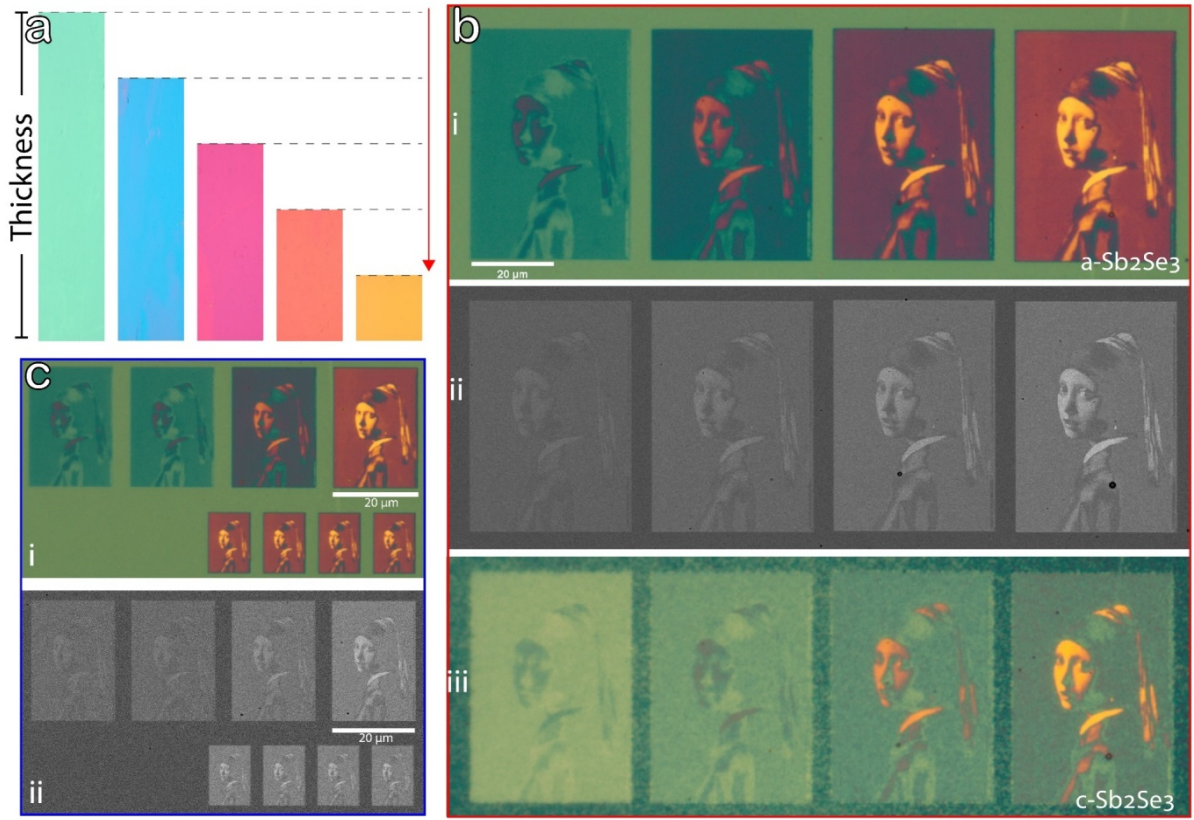

**Figure 5.** The added degree of freedom for super-resolution image formation at multiple depth levels and contrast formation at variable lateral dimensions. (a) Schematic description of the concept of depth modulation for contrast variation. Since the reflectance state depends highly on the film thickness, producing the metastructure heterostructure reflective states at varying depth levels makes different contrast states. (b) By linearly increasing the milling/sputtering power of the FIB, contrast can be produced at a specific location throughout the film thickness. In (i), four different contrast states are seen for the as-deposited phase of the $Sb_2Se_3$ film on gold. The relative height difference is the same for all contrast states, but the "contrast formation depth" is different. The BSE-SEM image in (ii) shows the differences in contrast since the distance to the gold differs for the four states. (iii) crystallization also induces another contrast state. In (c), the ability of the FIB milling to produce contrast in smaller lateral dimensions is shown. (i) The contrast modulation on a smaller lateral dimension (down to 10 μm width) is presented. (ii) the BSE-SEM images of the metastructure reflective device for varying lateral dimensions is presented.

An SEM image from the BSEs signal is presented in Fig. 5b(ii). The depth profile change is visualized in more detail from the contrast change of the produced nanostructures going from left to right. The structure from the left is placed close to the surface, and the BSEs signal is relatively low compared with the structure from the right. Milling/sputtering done in a relatively deeper region and thus close to the gold substrate have a more intense BSEs signal and appears brighter. Furthermore, the crystallization produced another set of contrasts from the optical parameter changes. Fig. 5b(iii) shows the structured surfaces after crystallization and makes another set of contrasts different from what was seen for the as-deposited amorphous phase of the $Sb_2Se_3$ thin film. Accessing all available depth levels throughout the film's thickness and the added degree of freedom from crystallization will immensely increase the functionality when moving toward a dynamic display device. In Fig. 4, we showed that nanoprinting digital images could achieve extremely high PPI values on a reflective device surface, but the PPI value can be even higher. Fig. 5c shows the production of the same structures in a series of smaller lateral dimensions. Executing designs in reduced dimensions offers the possibility of producing nanostructured surfaces with wide size tunability, which is crucial for future functionality. Additional examples of nanoprinting and depth-modulated contrast formation of the *Starry Night* by Vincent van Gogh are given in supplementary information (SI) Fig. S3.

In this work, we demonstrated the possibility of producing a super-high-resolution reflective nanostructured surface using FIB. The heterostructure device consists of a reflective gold substrate and a lossy PCM, $Sb_2Se_3$. By correlating the grayscale values of digital images to the milling/sputtering power of the FIB, local nanostructuring with sub-nm height variations control was achieved. Since small thickness changes of a thin film in the strong interference formalism translate to considerable reflectance changes, variable structural colors can be produced locally, thus creating contrast. Furthermore, the small probe size of the FIB is ideal for creating nanosized height contrasts in a relatively small lateral dimension creating vivid contrast with very high PPI values. Moreover, we showed our ability to prepare depth-modulated super-resolution nanostructure formation at any depth level throughout the thickness of the film, adding another degree of freedom for contrast tuning.

Compared with previously reported results for future dynamic display devices, using structured nanopillars[19,34] or through partial crystallization/amorphization of PCMs by electrical[16,18,23] or laser[15,17,20,22] signals for creating contrast, our ability to accurately reconfigure the height profile of neighboring 'pixels' with any lateral dimensions comes with obvious advantages. Creating the nanostructured by FIB is relatively easy, especially compared with classical methods like electron beam lithography (EBL), where creating local height variations is virtually impossible or requires multiple iterations. In contrast, this work shows that FIB can produce structures with relatively high lateral resolutions with extremely small pixel sizes. On the other hand, the ability of FIB to (1) create localized height variations and (2) access all depth profiles throughout the film's thickness to create contrasts is of significant importance when moving forward to a functioning dynamic display device. The above-mentioned abilities could only be achieved by FIB and are complex problems for other methods like EBL.

Another important point to note is the added value of the initially nanostructured surfaces to combine them in the future with known partial crystallization/amorphization techniques. In principle, one can produce a pixelated structured surface where neighboring pixels have different reflectance profiles, thus creating various colors (RGB, for example), mimicking traditional display devices and combinations that present different colors. Furthermore, this color mixing scheme could be achieved by partially tunning the reflectance of the individual pixel so that a dynamic display could be produced. Finally, at first sight, the current 'height-variations approach' might suggest that the structural colors would not work for grazing angle viewing. However, a closer look at the aspect ratio of nm scale height variations versus pixels of the order of a micron still makes it evident that the surface remains highly smooth and can still be observed well under grazing angle. On the contrary, the current approach using a strong interference effect is, in this respect, clearly advantageous compared to using Fabry-Perot type interference.

## Conclusions

This work demonstrated a local nanostructure of phase-change thin films to create structural colors and vivid contrasts for high-resolution dynamic image formation using a focused ion beam. Through a controlled milling/sputtering of materials, we showed we could create height differences with various lateral dimensions. Furthermore, we can generate structural colors locally by accessing the reflectance changes with the thickness of a heterostructure reflective device consisting of a phase-changing $Sb_2Se_3$ film on a reflective gold layer. We showed that structural color changes could produce a contrast with extremely small pixel sizes, producing high-resolution images with extremely high PPI values. Moreover, by linearly changing the milling power of the FIB, contrast can be modulated throughout the thickness of the film. This way, any available combinations of contrast levels and structural colors can be achieved. We believe our work will open up research on future dynamic display devices.

## Acknowledgments

This project has received funding from the European Union's Horizon 2020 Research and Innovation Programme "BeforeHand" (Boosting Performance of Phase Change Devices by Hetero- and Nanostructure Material Design)" under Grant Agreement No. 824957.

Supplementary Information for

# Structural colors and enhanced resolution at the nanoscale:

## Local structuring of phase change materials using focused ion beam

*Daniel T. Yimam\*, Minpeng Liang, Jianting Ye, Bart J. Kooi\**

Zernike Institute for Advanced Materials, University of Groningen, Nijenborgh 4, 9747 AG Groningen, The Netherlands

*Corresponding authors. Email: d.t.yimam@rug.nl, b.j.kooi@rug.nl

**SI 1 – Calibration digital image and thickness variation**

For high-accuracy controlled milling/sputtering, the focused ion beam (FIB) power vs. the amount of removed material has to be calibrated. This is done by correlating the grayscale value of a digital image with the final thickness of the produced structure for different milling powers. The main text discusses the calibration approach in detail in Fig. 3. Fig. S1a shows the original digital image used to calibrate the FIB power vs. thickness, with the produced structures in Fig. 3 of the main text. From the AFM image in Fig. 3c and the associated line profile in Fig. 3d, we can extract the depth variations of each square milled on the $Sb_2Se_3$ film surface. Fig. S1b shows the relationship between the grayscale values of the original digital image and the nanostructures' milling depth on the film's surface. A linear relation between the grayscale value and the milling depth can be seen. This information is critical for selecting appropriate milling parameters like accelerating voltage, beam current, and beam dwell time. Since the thickness range (from 0 – 26 nm, for example) has to be precisely correlated with the grayscale range ( 0 – 255), we can easily access any thickness range.

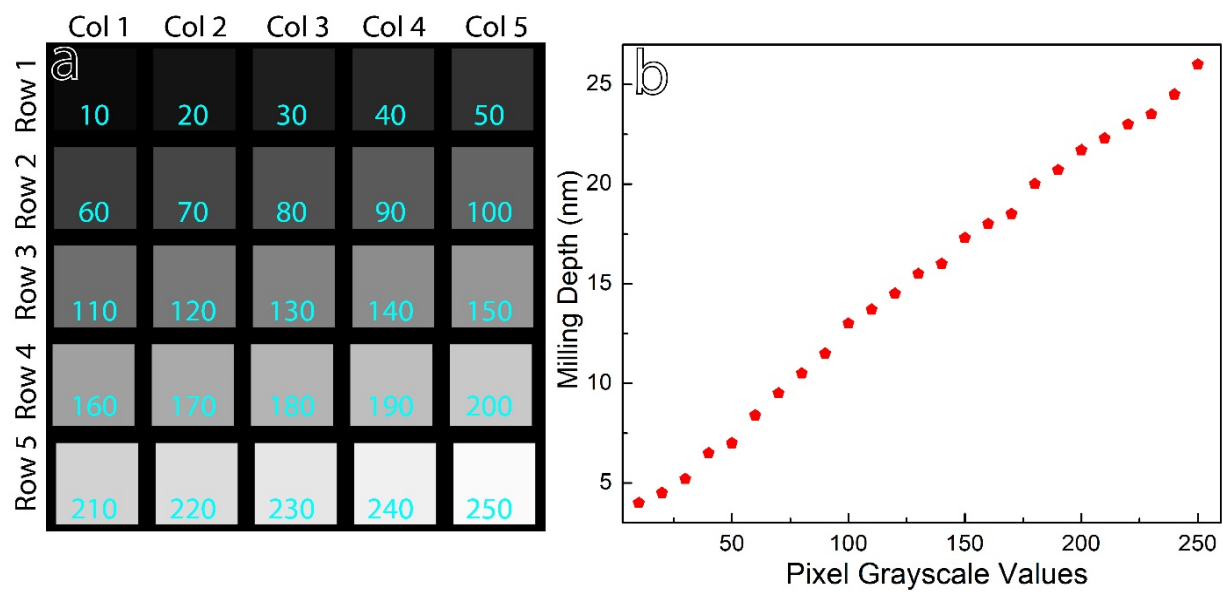

**Figure S1**. $Sb_2Se_3$ milling/sputtering calibration to the grayscale value of a digital image. (a) A digital image with 25 squares of varying 8-bit grayscale values from 10 to 250. The image was used to calibrate the milling/sputtering power of the focused ion beam (FIB) for different milling/sputtering parameters. (b) The milling/sputtering depth change with grayscale value. The relationship has been extracted from the AFM image and the associated line profiles shown in Fig. 3c and 3d of the main text.

## SI 2 – Reflectance calculated and measured

In Fig. S2, the calculated and measured reflectance profiles of locally produced structural colors (shown in Fig. 3e and 3f of the main text) are presented. The reflectance profiles are for the as-deposited amorphous ((a) and (c)) and the crystalline ((b) and (d)) phases of the $Sb_2Se_3$ film. The different structural colors produced different reflectance profiles corresponding to the thickness changes in the structured squares. The measured reflectance values for both phases of the $Sb_2Se_3$ film match very well with the calculated values.

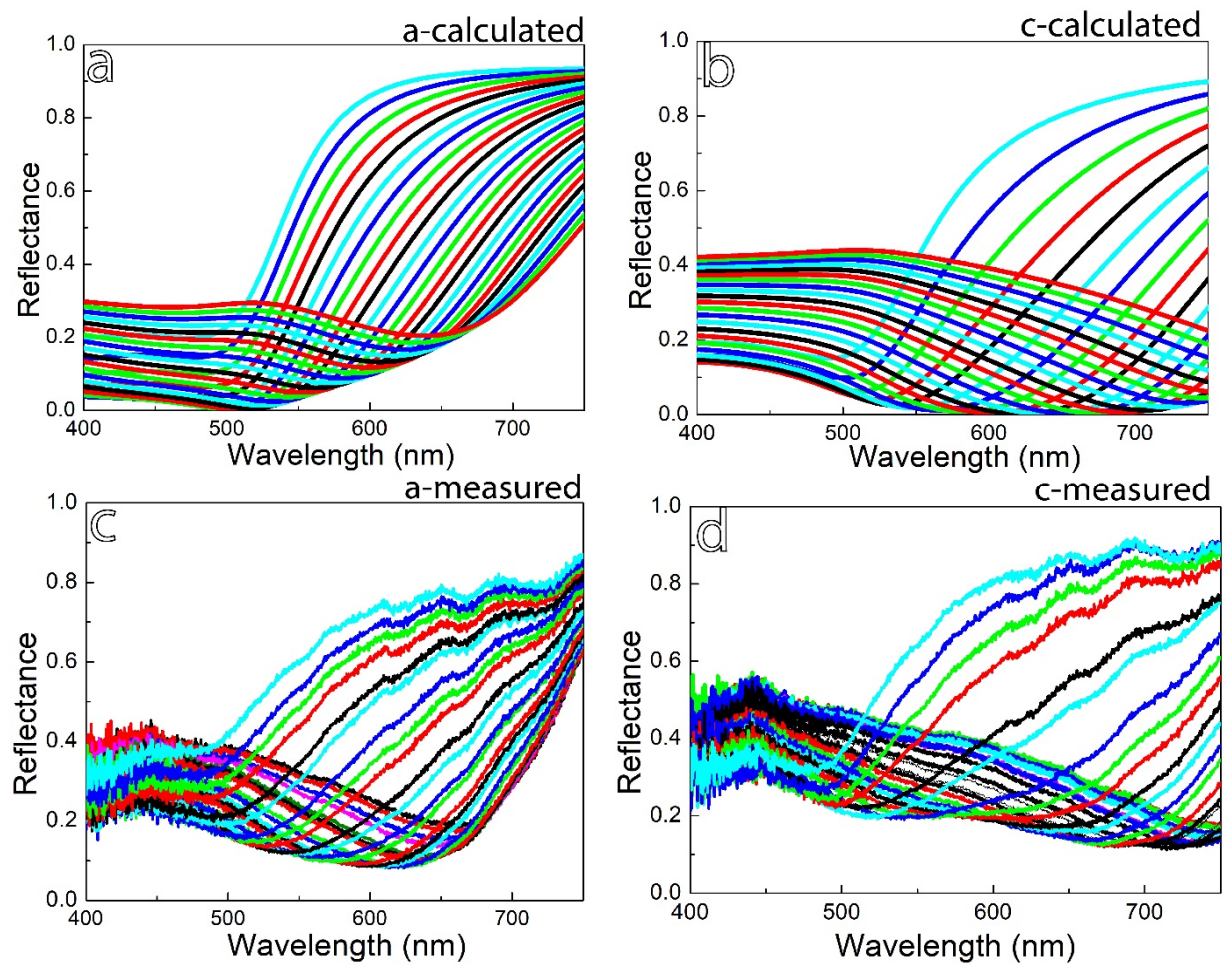

**Figure S2**. Calculated and measured reflectance profiles of locally created structural colors in Fig. 3e and 3f of the main text. In (a) and (b), the calculated reflectance curves for the as-deposited amorphous and crystalline phases of the $Sb_2Se_3$ layer are presented, respectively, for the reflective device shown in Fig. 3a of the main text. The $Sb_2Se_3$ layer thickness was changed from 3 – 26 nm to correlate the depth profile seen in Fig. S1b accurately. In (c) and (d), the measured reflectance profiles of the milled squares are presented for the as-deposited amorphous and crystalline phases, respectively. The measured reflectance profiles match very well with the calculated reflectance profiles.

**SI 3 – Contrast formation from depth modulation**

In Fig. 5 of the main text, the extended applicability of the FIB milling/sputtering technique used in this work to produce contrast at different depth levels throughout the thickness of the film is presented. An additional example, patterning *Starry Night* by Vincent van Gogh, is given here. As explained in detail in the main text, various contrasts can be produced by linearly increasing the milling/sputtering power of the FIB without compromising the relative height difference in each depth level.

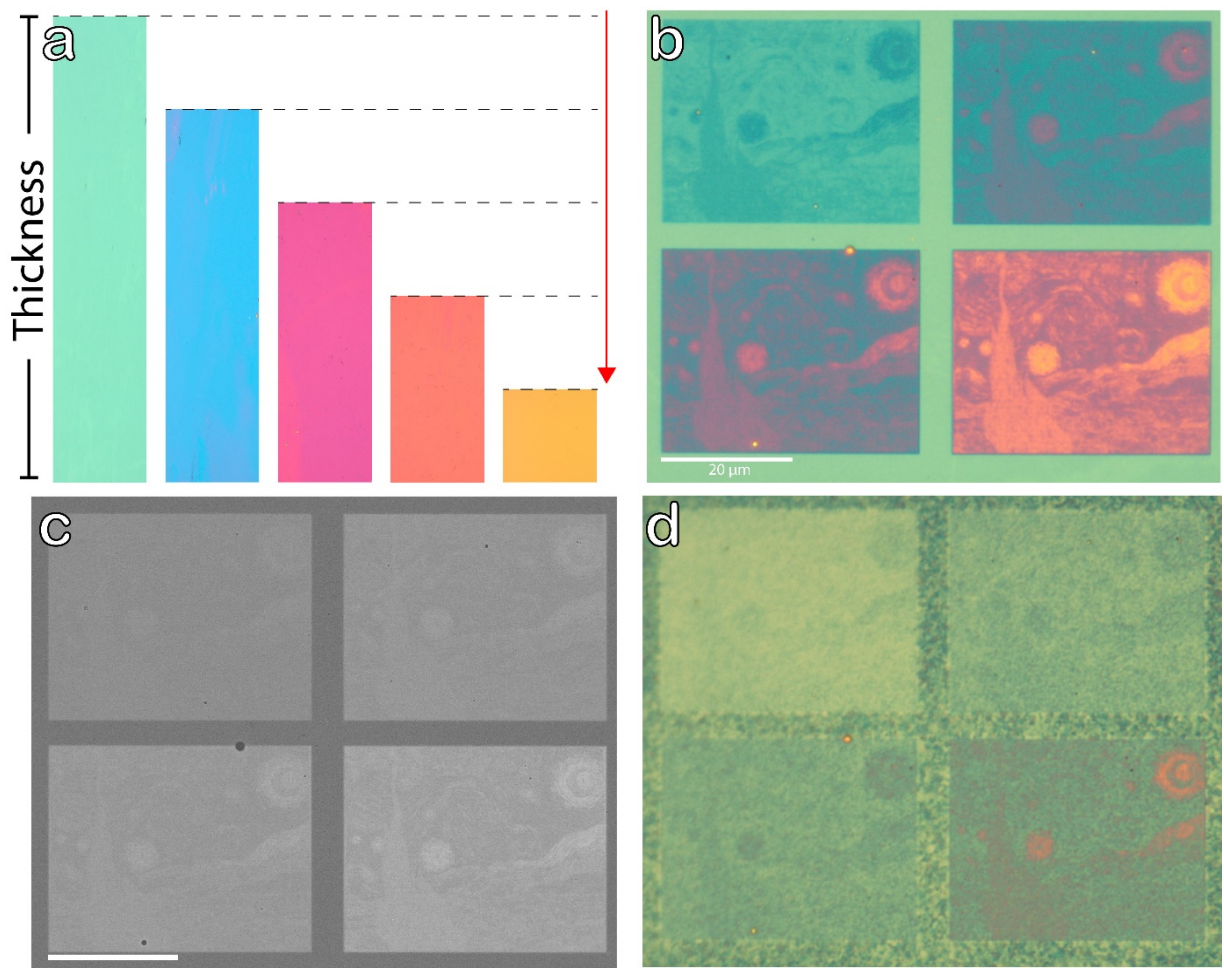

**Figure S3**. An additional example of contrast formation at multiple depth levels throughout the thin film thickness. The same principle used in Fig. 5 of the main text was used here to produce the contrasts. Here *Starry Night* by Vincent van Gogh was patterned on the film surface at different depth levels. (a) The schematic description of contrast creation at different depth levels. (b) An optical image is presented, showing multiple contrasts of the structured surfaces. The different contrasts are due to the difference in the depth level of the film's thickness. (c) The BSEs-SEM image shows a different contrast with an increasing signal when going to the deeper region of the film. (d) Crystallization also induces another set of contrasts for the produced structures.